\documentclass[twocolumn]{aastex6}
\usepackage{natbib}
\usepackage{graphicx}
\usepackage{epstopdf}
\usepackage{color}
\usepackage{listings}

\newcommand{\lya}{Ly$\alpha$}
\newcommand{\ha}{H$\alpha$}
\newcommand{\hb}{H$\beta$}
\newcommand{\oiii}{[OIII]$\lambda$5007}
\newcommand{\oiiit}{[OIII]$\lambda$4363}
\newcommand{\oii}{[OII]$\lambda$3727}

\def\arcsec{\hbox{$^{\prime\prime}$}}

\begin{document}
\title{Blueberry galaxies: the lowest mass young starbursts}
\author{Huan Yang\altaffilmark{1,2}, Sangeeta Malhotra\altaffilmark{2}, James E. Rhoads\altaffilmark{2}, Junxian Wang\altaffilmark{1} }
\altaffiltext{1}{CAS Key Laboratory for Research in Galaxies and Cosmology, Department of Astronomy, University of Science and Technology of China}
\altaffiltext{2}{Arizona State University, School of Earth and Space Exploration, huan.y@asu.edu}

\begin{abstract}
Searching for extreme emission line galaxies allows us to find low-mass metal-poor galaxies that are good analogs of high redshift \lya\ emitting galaxies. These low-mass extreme emission line galaxies are also potential Lyman-continuum leakers. Finding them at very low redshifts ($z\lesssim0.05$) allows us to be sensitive to even lower stellar masses and metallicities. 
We report on a sample of extreme emission line galaxies at $z\lesssim0.05$ (blueberry galaxies). We selected them from SDSS broadband images on the basis of their broad band colors, and studied their properties with MMT spectroscopy. From the whole SDSS DR12 photometric catalog, we found 51 photometric candidates. We spectroscopically confirm 40 as blueberry galaxies. (An additional 7 candidates are contaminants, and 4 remain without spectra.)  These blueberries are dwarf starburst galaxies with very small sizes ($< 1\hbox{kpc}$),  and very high ionization ([OIII]/[OII]$\sim10-60$). They also have some of the lowest stellar masses ($\log(\hbox{M}/\hbox{M}_{\odot})\sim6.5-7.5$) and lowest metallicities ($7.1<12+\log(\hbox{O/H})<7.8$) starburst galaxies. Thus they are small counterparts to green peas and high redshift \lya\ emitting galaxies. 
\end{abstract}

\section{Introduction}
One frontier in observational cosmology is understanding the reionization of the neutral hydrogen in the Universe. How did reionization occur and which sources caused it? Many simulations and observations suggest the very faint dwarf starburst galaxies (stellar mass $\sim 10^{6-8} M_{\odot}$) contribute a significant fraction of ionizing photons ({\it Lyman continuum, LyC}) during reionization (e.g. Trenti et al. 2010; Salvaterra et al. 2011; Bouwens et al. 2011; Dressler et al. 2015). Thus it is important to study these faint dwarf starbursts and understand their physical properties and formation mechanism.  

High redshift \lya\ emitters (LAEs) are an important population of low mass star-forming galaxies at $z > 2$, increasing in fraction to constitute 60\% of Lyman break galaxies at redshifts $z > 6$ (Stark et al. 2011). A large fraction of the dwarf starburst galaxies during reionization may be intrinsic LAEs, but their \lya\ photons can be scattered by the HI in IGM, which makes \lya\ line a powerful probe of reionization (e.g. Malhotra \& Rhoads 2004; Ouchi et al. 2010; Hu et al. 2010; Jensen et al 2013; Tilvi et al. 2014; Matthee et al. 2015; Zheng et al. 2017). 
These high-$z$ LAEs have low metallicity, low stellar masses, low dust extinction, and compact sizes (Gawiser et al. 2007; Pirzkal et al. 2007; Finkelstein et al. 2008; Pentiricci et al. 2009; Malhotra et al. 2012). 

Studying LAEs and faint dwarf starbursts in the high-$z$ universe is very challenging, because it usually requires long exposure times and difficult near-infrared observations (e.g. McLinden et al. 2011; Trainor et al. 2015). A complementary approach is to study the physical processes in low-$z$ analogs of these high-$z$ LAEs and faint dwarf starbursts. 

The current best nearby analogs of high-$z$ LAEs are green pea galaxies (e.g. Yang et al. 2016, 2017; Jaskot \& Oey 2014; Henry et al. 2015;  Verhamme et al. 2017). Green pea galaxies were discovered in the citizen science project Galaxy Zoo (Cardamone et al. 2009). They are compact galaxies that are unresolved in SDSS images. The green color is because the [OIII] doublet (EW(O[III]5007)$\sim$300-2500\AA) dominates the flux of SDSS $r$-band, which is mapped to the green channel in the SDSS's false-color {\it gri}-band images. They are similar to high-$z$ LAEs in many galactic properties -- small sizes, low stellar masses (10$^{8-10}M_{\odot}$), low metallicities for their stellar masses, high specific star formation rates (sSFR), and large \oiii/\oii\ ratios (Cardamone et al. 2009; Amorin et al. 2010; Izotov et al. 2011). Five green peas are observed and confirmed as LyC leakers (Izotov et al. 2016).

Green Peas are relatively luminous and massive galaxies compared to the faint-end dwarf starbursts and LAEs. In the present study, we searched for closer and lower mass analogs of the faint-end LAEs in SDSS $ugriz$ broadband images, and studied their properties with MMT and SDSS spectra. 

\begin{figure*}[htb!]
\centering
\includegraphics[width=0.9\textwidth]{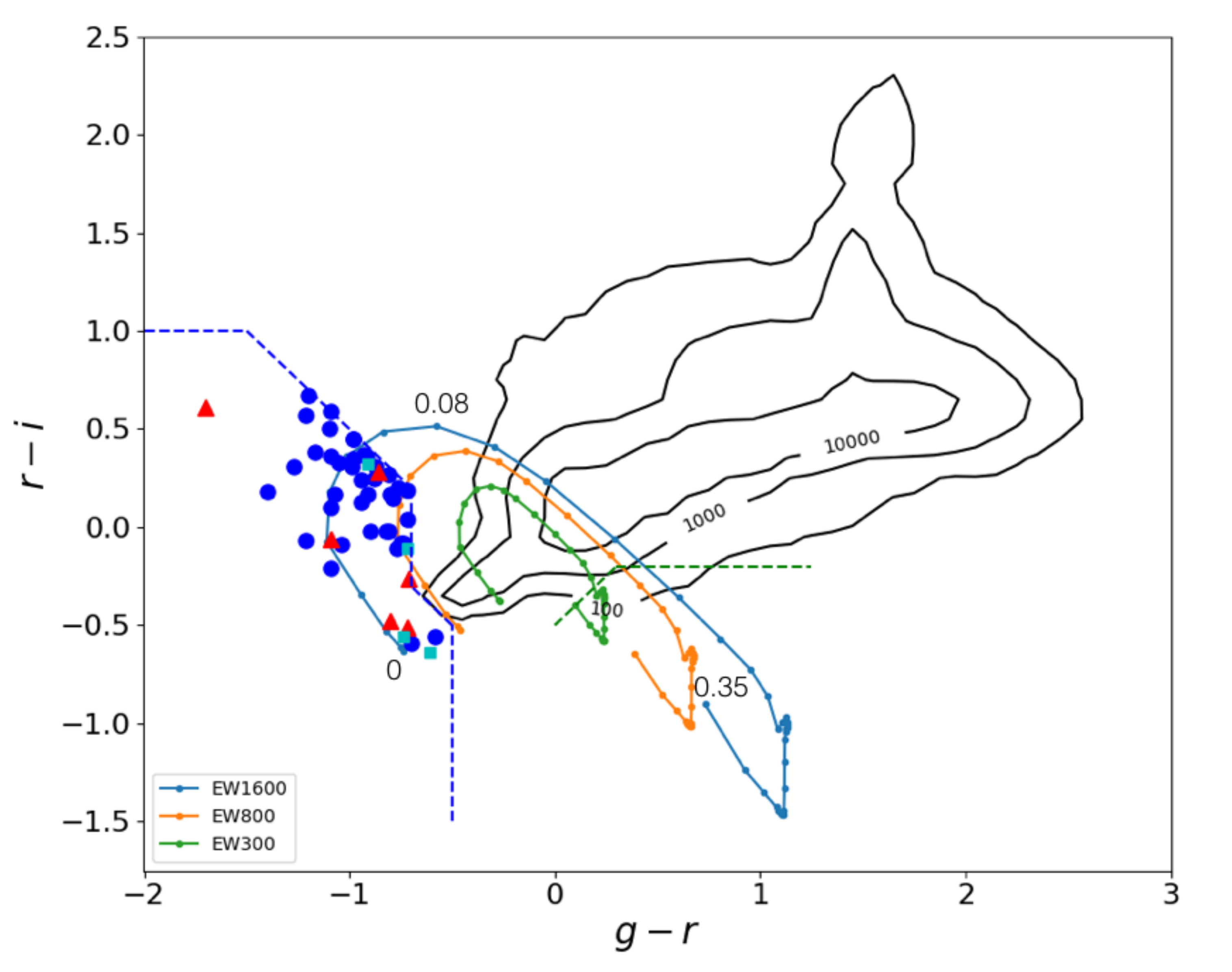}
\caption{The $g-r ~\hbox{vs.}~ r-i$ color-color diagram. The green, orange, and light blue solid lines with small dots are the redshift tracks for green peas with EW(\oiii)=300, 800, and 1600 \AA\ from $z=0$ to $z=0.35$. Each dot shows a step of 0.01 in redshift (redshifts 0, 0.08, and 0.35 are marked). Black contours are stars, normal galaxies, and quasars in the SDSS spectroscopic catalog with the numbers 100, 1000, and 10000 showing the source densities per 0.1$\times$0.1 mag grid. The green dashed line shows the green pea selection criteria in Cardamone et al. (2009). The blue dashed lines in the lower-left corner are the color selection criteria of blueberry galaxies with EW(\oiii)$\gtrsim800$\AA. A total of 51 sources are selected from the SDSS DR12 photometric catalog, including 40 blueberry galaxies confirmed with SDSS or MMT spectra (filled blue circles), 7 contaminants identified by MMT spectra (red triangles, of which one lies outside the plotted region), and 4 sources without spectra (cyan squares). These 40 blueberry galaxies are at $0.02<z<0.06$.
}
\end{figure*}

\begin{figure*}[ht]
\centering
\includegraphics[width=\textwidth]{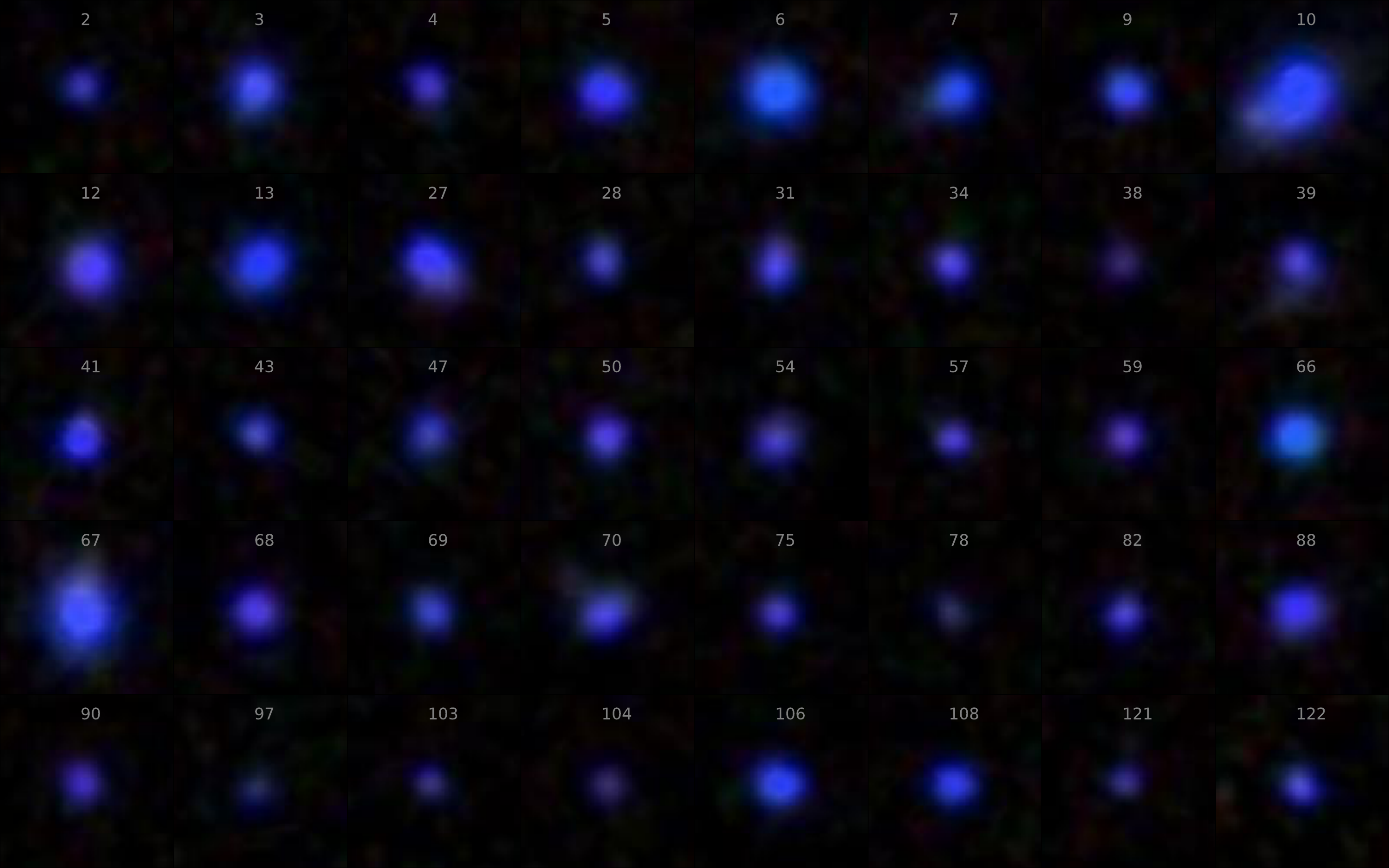}
\caption{The SDSS $gri$ images (Blue-Green-Red) of the 40 spectroscopically confirmed ``blueberry" galaxies. Each stamp is $10\times10\arcsec$. Even at $z\lesssim0.05$, many blueberry galaxies are unresolved or marginally resolved in SDSS images.
}
\end{figure*}

\begin{figure*}[ht]
\centering
\includegraphics[width=\textwidth]{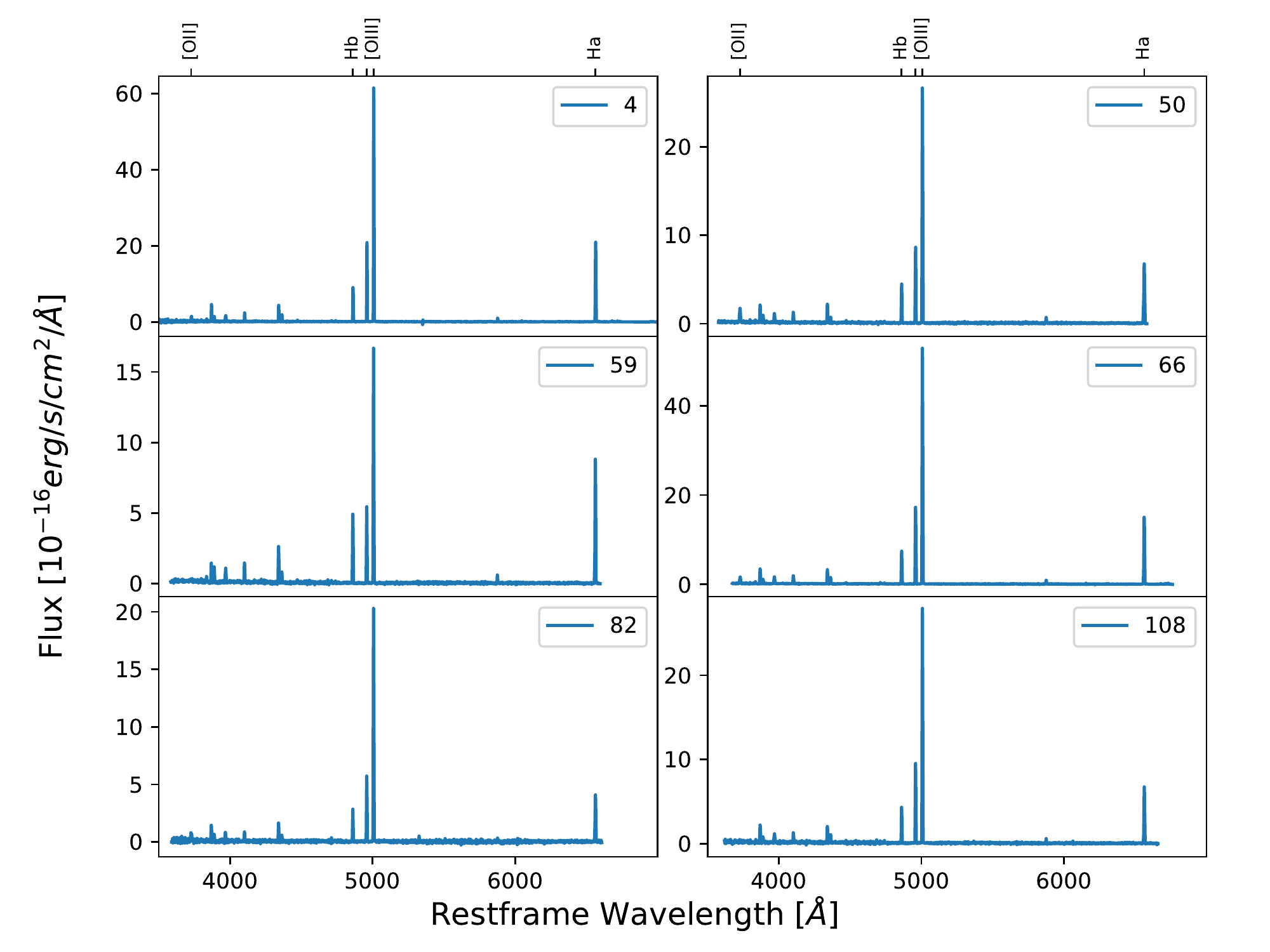}
\caption{The optical spectra of 6 representative blueberry galaxies. The emission lines of [OIII], [OII], \ha, and \hb\ are marked in the top two panels. The object ID is shown in each panel. The [OIII]5007 equivalent widths of this sample are about 700 -- 2400\AA.}
\end{figure*}

\begin{figure*}[ht]
\centering
\includegraphics[width=\textwidth]{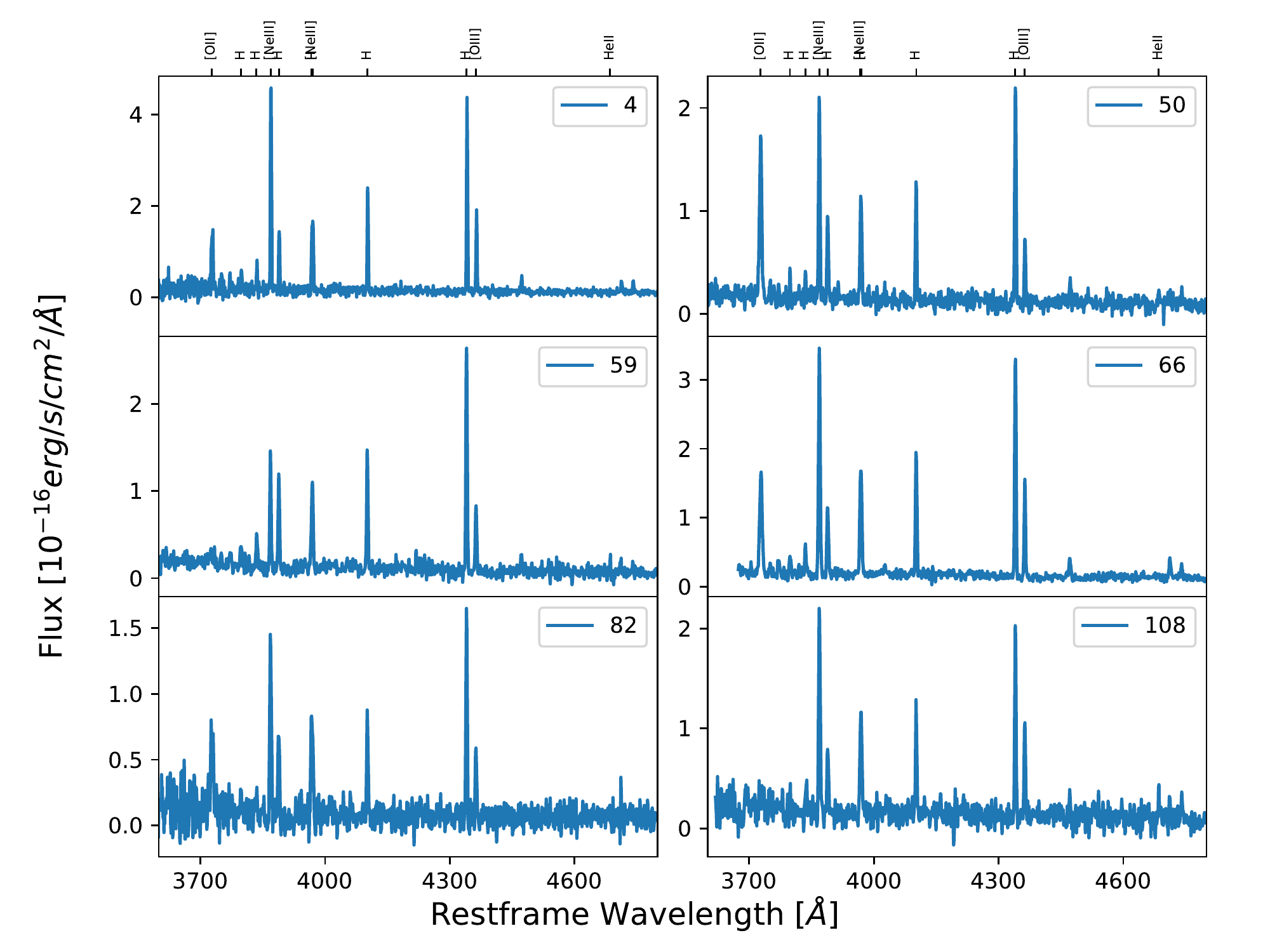}
\caption{The optical spectra in the 3600 -- 4800\AA\ range for the same 6 blueberry galaxies shown in fig.~3. In the upper two panels, we marked the emission lines of [OII]3727, [NeIII]3869, [NeIII3967, [OIII]4363, HeII4685, and a few H Balmer lines (H$\gamma$, H$\delta$, H$\epsilon$, H$\zeta$, H$\eta$, and H$\theta$). The [OIII]4363 line is very well detected. The object ID is shown in each panel.} 
\end{figure*}

\section{Photometric Selection of Blueberry Galaxies in SDSS}
The strong [OIII] emission line makes green peas show distinctive colors. Cardamone et al. (2009) selected green peas at $0.14<z<0.36$ in SDSS. To select the lowest mass green peas in SDSS images, we focus on the lowest redshift green pea sample. At the same limiting magnitudes of SDSS images, these lower redshift green peas generally tend to have lower luminosities and lower stellar mass. In this paper, we select green peas at $z\lesssim0.05$. 

We use simple color criteria to select photometric candidates of $z\lesssim0.05$ green peas. To get the color criteria, we made model spectra of green peas and generated corresponding redshift tracks in color-color space. The model spectra include both continuum and strong emission lines, with a range of line strengths. (1) The continuum component includes an age=4Myr young starburst and an  age=900Myr old population model spectra from Starburst99 (Leitherer et al. 1999). The old to young mass ratio is fixed at 10. (2) The strong emission lines include  [OII]3727, \hb, \ha, [OIII]4959, and [OIII]5007 at three different equivalent widths (EWs). The EW([OIII]5007) are 300, 800, and 1600 \AA. At each EW([OIII]5007), the strength of the other lines ([OII], \hb, \ha, and [OIII]4959) are estimated using typical line ratios of green peas at the corresponding [OIII]5007 equivalent width (Yang et al. 2017).  
Then we multiply the model spectra with the SDSS throughput curves for $ugriz$ filters and calculate the colors at different redshifts. When the \oiii\ and \ha\ lines move close to the wavelength edges of filters, the colors change dramatically with redshift.

According to the tracks in $g-r ~vs.~ r-i$ color-color diagram (figure 1), green peas at $z<0.08$ and $0.14<z<0.36$ have very different colors from stars, normal galaxies, and quasars. To select a clean sample of green peas at $z\lesssim0.05$, we use the following color selection criteria:
\[ g-r < -0.5 ~\hbox{and}~ r-i<1.0 ~\hbox{and}~ g-i<-0.5 \] 
\[~\hbox{and}~ (g-r<-0.7 ~\hbox{or}~ g-i<-1.0) ~\hbox{and}~ g-u<-0.3\]
The $g-u$ selection is added to exclude some very blue young stars or white dwarfs that fall in the selection region. Those stars have very blue $u-g$ colors in comparison to the red $u-g$ colors of green peas with strong \oiii\ in $g$ band. As shown in figure 1, these color criteria select green peas with EW(\oiii)$\gtrsim800$\AA\ at $z\lesssim0.05$. The selection was conducted with the SDSS DR12 photometric catalog. To exclude imaging artifacts from the selected catalog, we add some photometric flags to the selection. The exact selection criteria are in Appendix A. Especially, because green peas are compact sources, we use the flag ``NOT CHILD" to select small sources that are {\it not} deblended as a part of an underlying diffuse galaxy. 

Among all sources in the SDSS DR12 photometric catalog, only 51 sources satisfy the selection criteria. Figure 2 shows their SDSS $gri$ images (Blue-Green-Red jpg files generated by SDSS, Robert et al. 2004). Most of them are unresolved or marginally resolved in SDSS images (seeing $\sim1.3\arcsec$), although a few sources have some weak diffuse emission around the central bright knots. Because their blue/purple colors, we call them ``blueberry" galaxies. 

The Pan-STARRS1 (PS1) survey (Chambers et al. 2017) has deep $z$ and $y$ band images, so we also get the PS1 $grizy$ photometric data for this sample to aid in spectral energy distribution (SED) fitting. The photometric data are in Table 1 and in the attached machine-readable table.

\section{Optical Spectra of Blueberry Galaxies}
We obtained spectra of this sample from SDSS, MMT, and LBT. Out of the total sample of 51 sources, 10 sources already have SDSS spectra. These 10 sources (the first 10 objects in figure~2, with objID $<15$) all show strong [OIII] emission lines and are at the bright end of this sample. One object was observed using LBT/MODS (5 minutes exposure) in the morning of May 07, 2016, after twilight started. The spectra shows  strong emission lines. Another 36 objects were observed with MMT/BlueChannel on Jan 07, 2017. The remaining 4 objects don't have optical spectra.  In all spectra, the wavelength range of [OII]3727 to \ha\ was covered. Out of the 47 objects with spectra, 40 objects show extreme emission lines. The other 7 objects are stars, quasars, or unknown sources (with low S/N spectra).

The MMT observations were made using the medium-resolution grating (500GPM) and 1.25\arcsec $\times$ 180\arcsec\ slit, giving a spatial scale along the slit of 0.6\arcsec\ per pixel, a spectral range of 3700-6900 \AA, and a spectral resolution of $\sim$3.8 \AA\ (FWHM). 
The seeing during the MMT observations was $\sim0.7-1.0 \arcsec$. The slit orientations were along the parallactic angles for all objects. The exposure time is about 10 -- 15 minutes for each object. 

The MMT/BlueChannel and LBT/MODS long-slit spectra were reduced using IRAF following the standard steps. The raw data were bias subtracted and flat-fielded. The sky background at the upper and lower regions of the object were subtracted. The continuum was detected in most objects and was used for determining the spectral trace. The traces of two faint objects were determined using other spectra as references. The 1-D spectra were extracted using optimal extraction. The spectra of HeNeAr lamps were used for wavelength calibration. The standard star LB227 was observed with 5\arcsec\ slit width and was used for the flux calibration. To get better absolute spectral fluxes, we calculated a $g$-band magnitude by multiplying the spectra with $g$ band throughput curve, and then scaled each spectrum to make the spectroscopic $g$ magnitude equal the SDSS $g$ band magnitude. 

The optical spectra (figure 3) of blueberry galaxies show very strong [OIII]5007 emission lines, and large [OIII]5007/\hb\ and [OIII]/[OII] ratios. Figure 4 shows the spectra in 3600 -- 4800\AA. The [OIII]4363 line is very strong and well detected. The [OII]3727 line is relatively weak. 

From the optical spectra, we measured the properties of emission lines. Firstly, we estimate the continuum by fitting a cubic polynomial function to wavelength regions without emission lines and subtract the continuum from the spectra. Then we fit the emission lines with gaussian functions and get the line fluxes. We correct the measured line fluxes for Milky Way extinction using the attenuation of Schlafly \& Finkbeiner (2011) (obtained from the NASA/IPAC Galactic Dust Reddening and Extinction tool) and the Fitzpatrick (1999) extinction law. Notice that the \ha/\hb\ ratios for most of the MMT spectra are smaller than the case-B value 2.86, because the flat field calibration in the red end of spectra ($6000-6900\AA$) is very poor and the \ha\ fluxes are probably underestimated. Therefore we didn't correct the line fluxes for extinction by dust within the blueberry galaxies. The properties of emission lines are shown in Table 2 and more columns are included in the attached machine-readable table.

\begin{figure}[ht]
\centering
\includegraphics[width=0.5\textwidth]{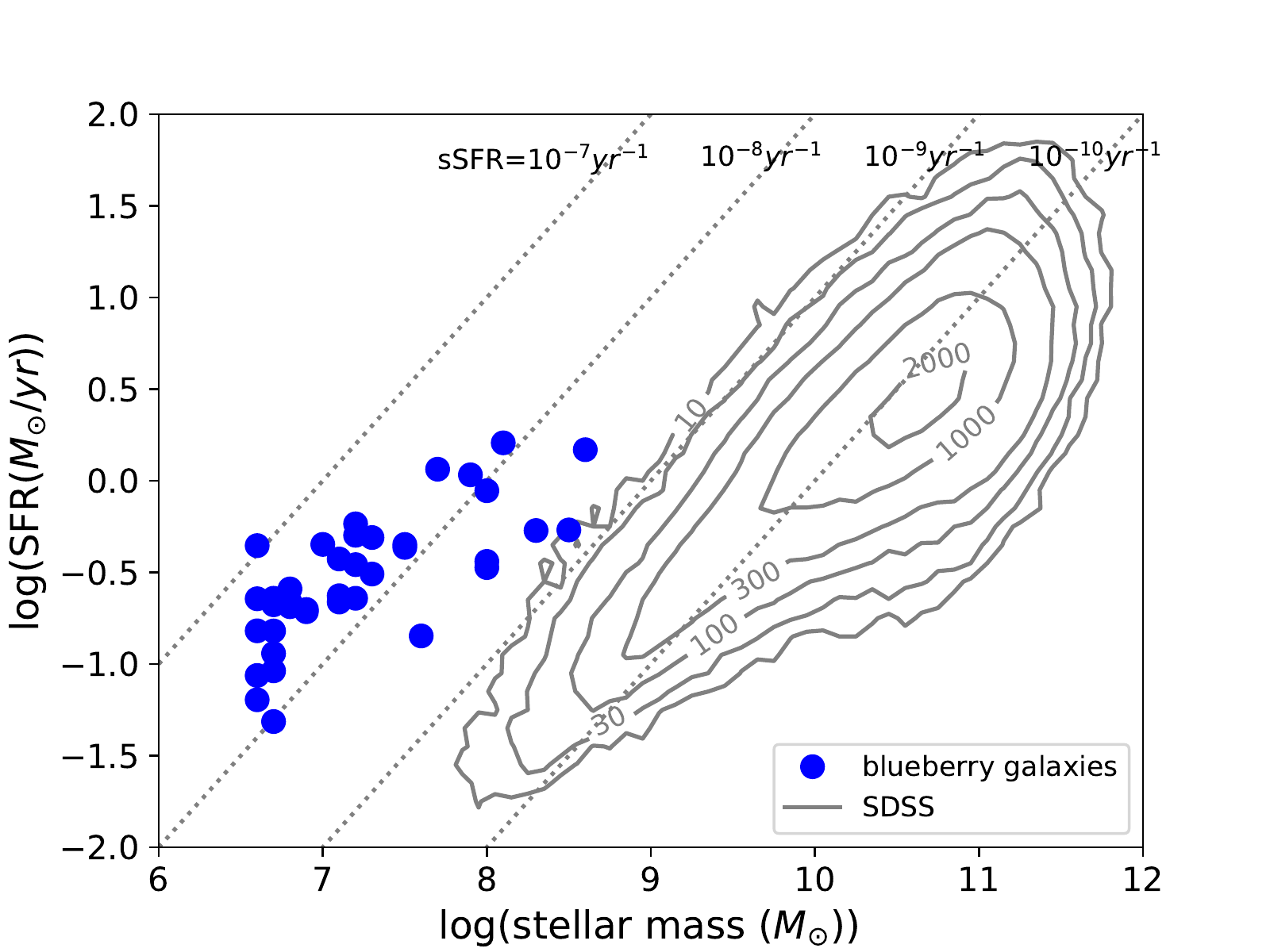}
\caption{The stellar mass vs. star formation rate (SFR) for blueberry galaxies. The dotted lines show constant sSFR of $10^{-7}~yr^{-1}$, $10^{-8}~yr^{-1}$, $10^{-9}~yr^{-1}$, and $10^{-10}~yr^{-1}$. The grey contours show the mass$-$SFR relation of $\sim$700,000 SDSS galaxies with redshifts at $0.01<z<0.30$ and magnitude$<$18 (Salim et al. 2016).} 
\end{figure}

\begin{figure}[ht]
\centering
\includegraphics[width=0.5\textwidth]{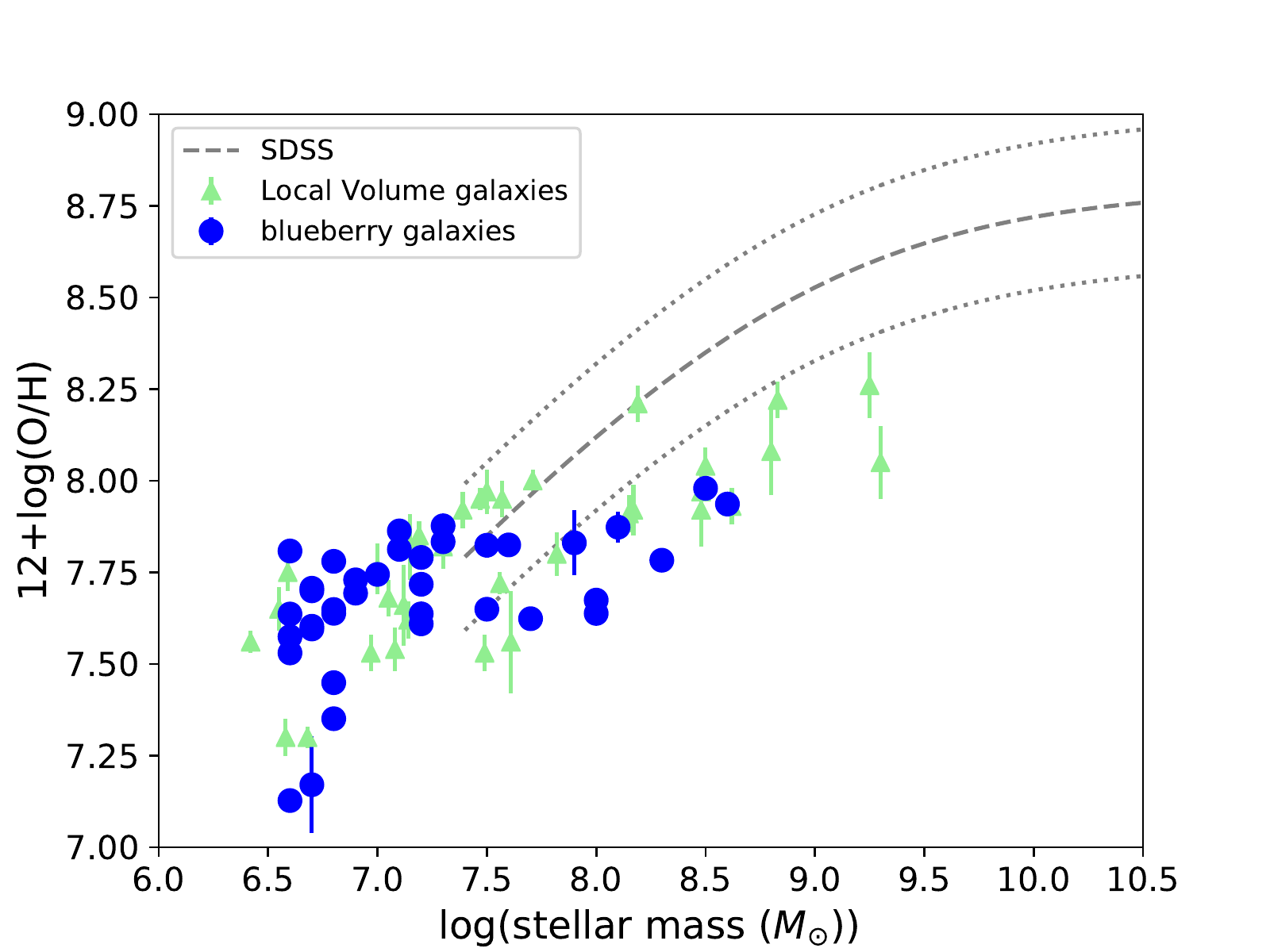}
\caption{The stellar mass vs. gas metallicity of blueberry galaxies (blue dots; the metallicity errorbar is smaller than the blue dot for many galaxies). The dashed line shows the mass--metallicity relation of SDSS galaxies measured from stacked spectra in Andrews \& Martini (2013) and the dotted lines show its 1$\sigma$ scatter ($\sim$0.2dex). The green triangles are galaxies in the Local Volume Legacy survey (Berg et al. 2012; Lee et al. 2006). The metallicities of blueberries, local volume galaxies, and SDSS galaxies are all calculated using the [OIII]4363 line.} 
\end{figure}

\begin{figure}[ht]
\centering
\includegraphics[width=0.5\textwidth]{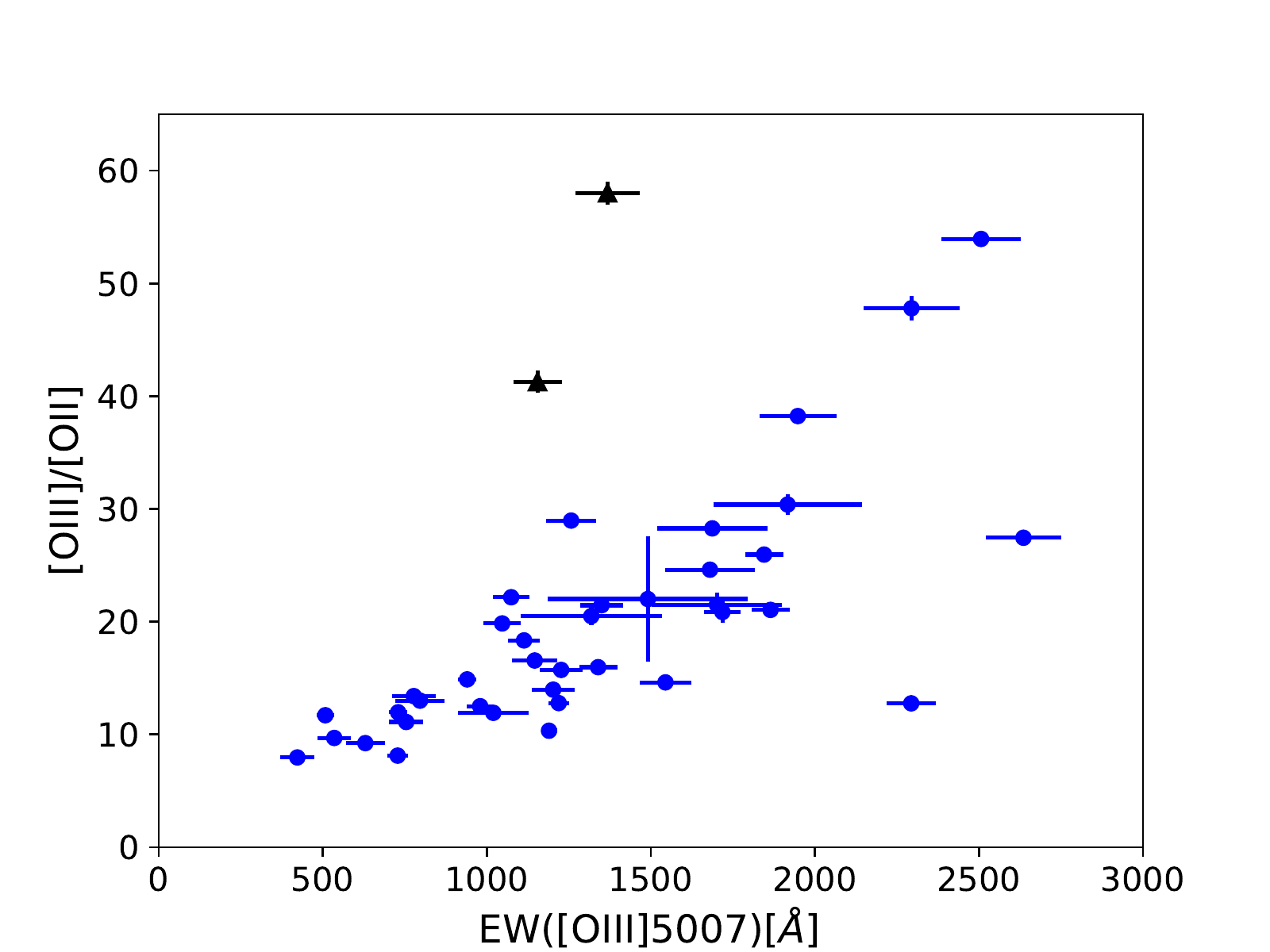}
\caption{The [OIII]/[OII] vs. EW([OIII]5007) of blueberry galaxies. In two galaxies  (black triangles) where the [OII] line is detected at 2$\sigma$ level, we use the 3$\sigma$ upper-limit of [OII] line to calculate the [OIII]/[OII] ratio.} 
\end{figure}

\section{What are blueberry galaxies?}
From the optical spectra, we measured some physical properties of blueberry galaxies. These properties are shown in figures 5--8. \\ 
{\bf Stellar mass}: The stellar mass is estimated by fitting Starburst99 models (Leitherer et al. 1999) to the SDSS $ugri$ and PS1 $zy$ bands photometric data. The strong emission lines are subtracted from the SDSS $gri$ band photometry. We model the star formation history with two starburst components -- one young component with age$<20Myr$, and one old component with age$<14Gyr$. From the Starburst99 single stellar population model (generated with Kroupa initial mass function and Geneva 2012/2013 tracks with zero rotation at metallicity Z=0.002), we make 12600 composite spectra at a grid of three different parameters -- the age of young component, the age of old component, and the mass ratio of old to young components. Then we fit each composite spectrum to the SED and compute the likelihood of each spectrum as $\exp(-\chi^{2}/2)$. From the likelihood distribution, we estimate the median value (50\% probability on each side) as the final stellar masses and ages. The majority of the blueberry galaxies have stellar mass about 10$^{6.5}M_\odot$--$10^{7.5}M_{\odot}$. 

Note that a faint diffuse stellar disk may exist around the blueberry galaxies. The SDSS r-band images have a surface brightness limit of $\sim$26 $mag~arcsec^{-2}$, which corresponds to $\sim~10^{6}~L_{\odot}~kpc^{-2}$. Assuming a mass to light ratio $M_{r}/L_{r}=1$ for low mass galaxies (Kauffmann et al. 2003), the diffuse stellar disk has a surface limit of $\sim~10^{6}~M_{\odot}~kpc^{-2}$.

{\bf Star formation rate}: Assuming the Case-B condition, we convert the \hb\ luminosity to SFR using the formula $SFR(M_{\odot}/\hbox{yr})=L_{H\beta}(\hbox{erg}/\hbox{s})\times2.86\times10^{-41.27}$ (Kennicutt \& Evans 2012). (We estimate star formation using \hb\ rather than \ha\ because our spectra are better calibrated at \hb.) The resulting star formation rates are about $\hbox{SFR} \approx 0.05-2~M\odot~\hbox{yr}^{-1}$, and sSFR are about $10^{-8}~yr^{-1}$. Figure 5 shows the mass$-$SFR relation of blueberry galaxies. 

{\bf Gas metallicity:} We calculate the metallicity using \oiiit, \oiii, and \oii\ line fluxes following the $T_{e}$ method described in Izotov et al. (2006) and Ly et al. (2014). Blueberry galaxies show very low metallicities ($7.1<12+log(O/H)<8.0$). A few can be classified as extremely metal poor galaxies (XMP). And two blueberry galaxies have $12+log(O/H)<7.2$, and are among the most metal-poor galaxies ever found (e.g. Guseva et al. 2015). Thus, searching for blueberry galaxies may be a good method to find XMP with the lowest gas metallicity. Figure 6 shows the mass-metallicity relation of blueberry galaxies.

{\bf Ionization}: We also calculate the [OIII]/[OII] ratios (([OIII]4959 + [OIII5007) / [OII]3727) (figure 7). For two galaxies where the [OII] line is detected at $\la 2\sigma$ significance, we use the 3$\sigma$ upper-limit of [OII] line to calculate the [OIII]/[OII] ratio. The [OIII]/[OII] ratios for the full sample range from $\sim 8$ to $\ga 60$.

\begin{figure}[ht]
\centering
\includegraphics[width=0.5\textwidth]{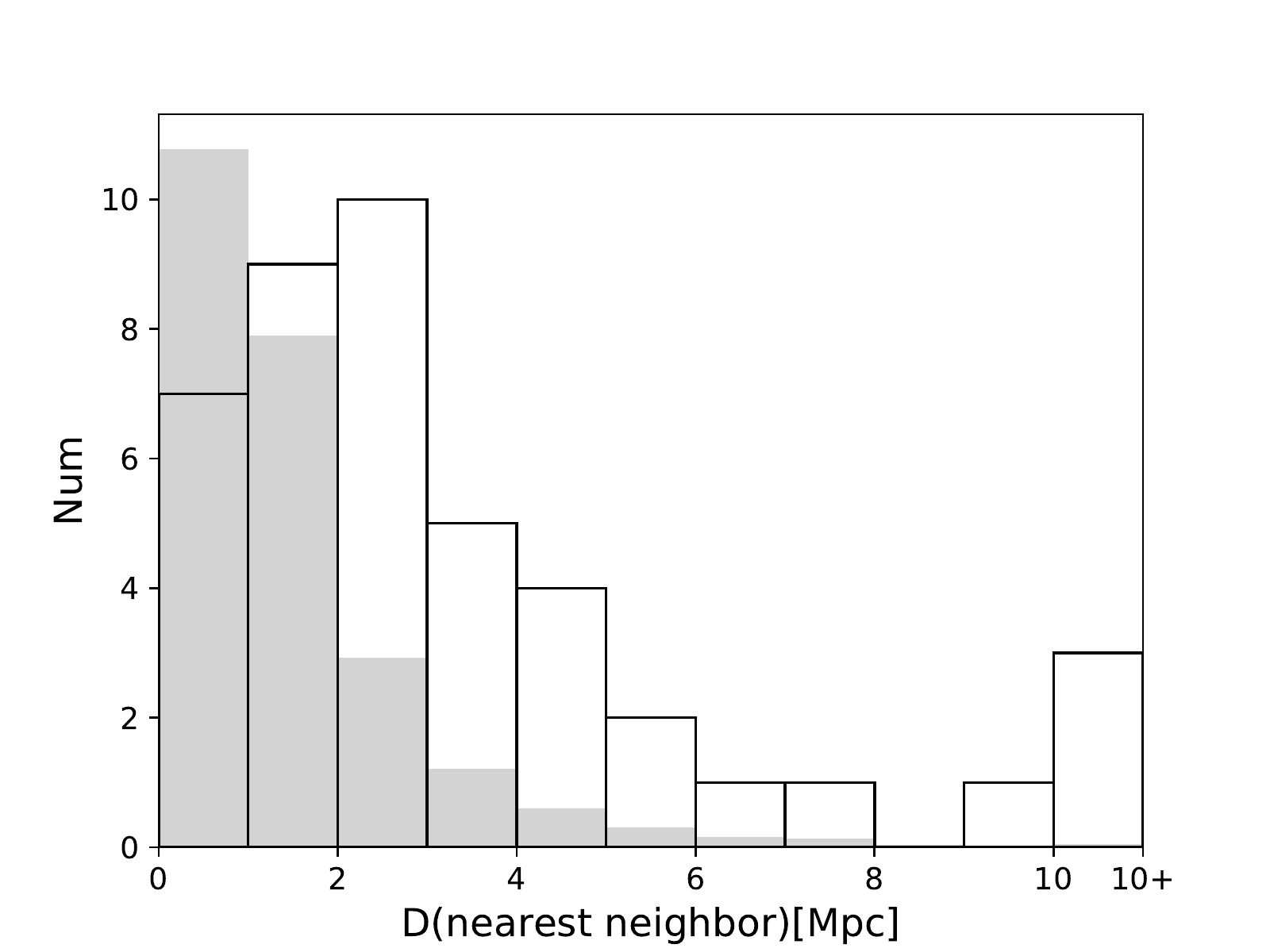}
\caption{The nearest neighbor distances of blueberry galaxies. The hollow black histogram shows the distribution of 3D distances from blueberry galaxies to their nearest neighbor galaxies in the SDSS DR12 spectroscopic catalog. The last bin includes 3 galaxies with D$>$10Mpc. The grey shaded normalized histogram shows the 3D distances to the nearest galaxy neighbor for 5000  galaxies at similar redshift ($0.03<z<0.05$) randomly selected from the SDSS DR12 spectroscopic catalog.} 
\end{figure}

{\bf Environment}: To characterize the environments of blueberry galaxies, we find the nearest galaxy neighbor in the SDSS DR12 spectroscopic catalog for each one. The distances to the nearest neighbor are between 300kpc $-$ $>$10Mpc (figure 8). We also calculate the nearest neighbors for a comparison sample of 5000 galaxies randomly selected at similar redshift ($0.03<z<0.05$) in the SDSS DR12 spectroscopic catalog. Most galaxies in the comparison sample have neighbors within 2Mpc. Blueberry galaxies and the comparison sample have different distributions of distances to nearest neighbor (with K-S test p-value=$10^{-39}$). So blueberry galaxies are generally in low density environments, and some are in the outskirts of galaxy groups. 

{\bf Number density}: Among all sources in the SDSS DR12 photometric catalog, only 51 sources satisfy the selection criteria (40 blueberries, 7 contaminants, 4 no spectra). Assuming 3 objects without spectra are blueberries, we get 43 objects in a sky area of 14555 deg$^{2}$ at redshift$<0.05$. Thus the number density of blueberry galaxies is 43/ (14355 / 41259 $\times$ 0.04 Gpc$^{3}$) = 3.0$\times$10$^{-6}$ Mpc$^{-3}$. 

In summary, these blueberry galaxies are currently undergoing a young starburst. They have very small sizes ($< 1kpc$), very low stellar masses and metallicities, and very high ionization. 

Compared to the many different classes of star-forming ``dwarf" galaxies, are blueberry galaxies a new class? The star-forming dwarf galaxies include the local volume dwarf galaxies (e.g. Lee et al. 2009), the blue compact dwarf galaxies (BCDs, e.g. Zwicky \&\ Zwicky 1971; Thuan \&\ Martin 1981; Gil de Paz et al. 2003), ultracompact blue dwarf galaxies (e.g. Corbin et al. 2006), HII galaxies (e.g. Terlevich \& Melnick 1981; Melnick et al. 2017), extreme metal-poor galaxies (e.g. Guseva et al. 2015; Sanchez Almeida et al. 2016), HI selected dwarfs (e.g. Huang et al. 2012), emission line dots (e.g. Werk et al. 2010; Kellar et al. 2012), and blue diffuse galaxies (James et al. 2017). Compared to those star-forming dwarf galaxies, blueberry galaxies have similar stellar mass and luminosity, but much stronger [OIII] line strength and gas ionization. Because blueberry galaxies are selected by the strong [OIII] emission lines, they represent the star-forming dwarf galaxies with the highest emission line strength and gas ionization. 

On the other hand, compared to green pea galaxies at $z\sim0.2-0.3$ and typical high-$z$ LAEs found in the current narrow-band surveys, blueberry galaxies have similarly strong emission lines but about 10-100 times smaller stellar mass, SFR, and luminosity. So blueberry galaxies represent the faint-end of green peas and LAEs.

\section{Conclusion}
We searched for blueberry galaxies in SDSS broadband images and studied their properties with MMT and SDSS spectra.  Our main results are as follows. 

(1) Using our color selection criteria, we find 51 photometric candidates at $z\lesssim0.05$ from the whole SDSS DR12 photometric catalog.  Optical spectra confirm that 40 of these sources show strong emission lines and can be classified as blueberry galaxies.  The remaining 11 candidates consist of 7 contaminants and 4 without spectra.  The number density of blueberry galaxy is about 3.0$\times$10$^{-6}$ Mpc$^{-3}$. 

(2) These blueberries are dwarf starburst galaxies with very small sizes ($<1\hbox{kpc}$), low stellar masses ($\log(M/M_{\odot})\sim6.5-7.5$), small SFR ($0.05-2~M_\odot\,\hbox{yr}^{-1}$),  high sSFR ($3-100~\hbox{Gyr}^{-1}$), very low gas metallicities ($7.1<12+\log(O/H)<8.0$) estimated with the [OIII]4363 lines, and very high gas ionization ([OIII]/[OII]]$\sim10-60$). Two blueberry galaxies have $12+\log(O/H)<7.2$ and are among the most metal-poor galaxies known. Their nearest-neighbor distances in the SDSS spectroscopic catalog are between 300kpc$-$10Mpc, larger than the galaxy population in general. Thus, they are in low density environments. These blueberry galaxies represent the faint-end sample of green peas and LAEs.

\begin{deluxetable*}{lrrlcccccccc}
\tablecaption{The sample}
\tabletypesize{\scriptsize}
\centering
\tablehead{\colhead{objID} & \colhead{RA} & \colhead{DEC} & \colhead{redshift} &  \colhead{$u$} & \colhead{$g$} & \colhead{$r$}  &\colhead{$i$} & \colhead{$z$} & \colhead{$y$} & \colhead{EW(\oiii)[\AA]}  &\colhead{D$_{neighbor}$[Mpc]}\\
\colhead{(1)} & \colhead{(2)} & \colhead{(3)} & \colhead{(4)} & \colhead{(5)} & \colhead{(6)} & \colhead{(7)} & \colhead{(8)}  & \colhead{(9)} & \colhead{(10)} & \colhead{(11)} & \colhead{(12)} } 
\startdata
  2 & 125.08029 & 54.52780 & 0.03858 &  21.40 & 20.46 & 21.55 & 21.45 &   21.56 &   -    & 1687 &  3.91 \\
  3 & 126.41854 & 18.77145 & 0.03792 &  19.81 & 19.02 & 19.83 & 19.85 &   20.19 &   20.30 & 1339 &  0.61 \\
  4 & 141.73101 & 45.07562 & 0.04225 &  21.25 & 20.11 & 21.11 & 20.88 &   21.03 &   -    & 1947 &  4.23 \\
  5 & 158.23637 & 49.32979 & 0.04403 &  19.93 & 18.74 & 20.01 & 19.70 &   20.24 &   20.29 & 2635 &  0.54 \\
  6 & 200.94776 &  -1.54778 & 0.02245 &  19.27 & 18.12 & 18.82 & 19.41 &   19.74 &   19.43 & 2055 &  2.26 \\
  7 & 208.85694 & 46.86427 & 0.02811 &  20.38 & 19.31 & 20.06 & 20.14 &   20.82 &   20.56 & 1318 &  1.17 \\
  9 & 221.17238 &  4.16159 & 0.03875 &  20.51 & 19.41 & 20.31 & 20.33 &   20.45 &   20.50 & 1349 &  7.17 \\
 10 & 227.39239 & 37.52948 & 0.03259 &  18.36 & 17.32 & 18.09 & 18.20 &   18.51 &   18.38 & 1718 &  2.22 \\
 12 & 239.10198 & 48.11272 & 0.05024 &  19.81 & 18.75 & 19.73 & 19.38 &   19.86 &   19.78 & 1219 &  2.57 \\
 13 & 242.04318 & 35.46926 & 0.03274 &  19.94 & 18.67 & 19.76 & 19.97 &   20.20 &   20.06 & 2294 &  2.45 \\
 27 &  26.72211 &  3.32288 & 0.04672 &  19.65 & 18.65 & 19.70 & 19.37 &   19.94 &   19.95 & 1226 &  9.24 \\
 28 &  34.09903 & 17.25720 & 0.03921 &  20.95 & 20.18 & 20.97 & 20.82 &   21.38 &   21.13 &  796 & 10.41 \\
 31 &  39.70384 &  1.40929 & 0.04983 &  20.50 & 19.64 & 20.56 & 20.20 &   20.47 &   -    &  754 &  6.05 \\
 34 &  59.30632 & 18.14599 & 0.03732 &  21.85 & 19.98 & 20.91 & 20.54 &   20.93 &   21.02 & 1916 & 25.34 \\
 38 & 121.11169 & 40.36818 & 0.05768 &   -    & 20.92 & 22.01 & 21.42 &   21.47 &   -    & 1074 &  2.71 \\
 39 & 126.94439 & 10.98644 & 0.04355 &  20.97 & 19.72 & 20.53 & 20.26 &   20.98 &   20.54 & 1046 &  0.89 \\
 41 & 129.40511 & 18.39092 & 0.04094 &  20.41 & 19.06 & 20.46 & 20.28 &   20.66 &   20.64 & 2506 &  2.32 \\
 43 & 133.40643 & 23.49403 & 0.04310 &  20.91 & 20.22 & 21.29 & 21.12 &   21.33 &   -    &  729 &  2.25 \\
 47 & 140.59943 & 63.41027 & 0.03931 &  21.13 & 20.03 & 20.94 & 20.77 &   21.18 &   21.02 &  979 &  1.69 \\
 50 & 150.58734 & 31.49262 & 0.05134 &  20.64 & 19.75 & 20.84 & 20.48 &   20.93 &    -    & 1202 &  2.43 \\
 54 & 156.65863 &  4.44845 & 0.04215 &  20.73 & 19.93 & 20.80 & 20.49 &   21.06 &   20.42 &  728 &  0.30 \\
 57 & 158.87949 & 14.00532 & 0.03970 &  21.46 & 20.36 & 21.12 & 20.92 &   20.96 &   -    & 1146 &  4.85 \\
 59 & 161.53849 & 40.78530 & 0.04897 &  20.85 & 20.16 & 21.36 & 20.69 &   21.74 &   -    & 1154 &  3.36 \\
 66 & 168.30101 &  3.02023 & 0.02336 &  19.91 & 18.83 & 19.41 & 19.97 &   20.22 &   19.86 & 1845 &  1.41 \\
 67 & 170.95395 & 20.84203 & 0.03283 &  18.71 & 17.57 & 18.39 & 18.41 &   18.70 &   18.34 & 2293 &  0.88 \\
 68 & 173.24695 &  8.16188 & 0.04944 &  20.73 & 19.51 & 20.72 & 20.15 &   20.99 &    -    & 1680 &  1.65 \\
 69 & 174.22574 & 34.45777 & 0.03492 &  20.97 & 20.05 & 20.79 & 20.87 &   21.28 &    -    &  -  &  1.82 \\
 70 & 174.75171 &  0.67850 & 0.04165 &  20.25 & 19.45 & 20.25 & 20.08 &   20.37 &   19.86 &  940 &  2.98 \\
 75 & 184.59971 & 39.41910 & 0.05093 &  21.20 & 20.29 & 21.28 & 20.97 &   -    &   -    & 1113 &  5.15 \\
 78 & 195.85314 & 39.48141 & 0.04779 &  22.22 & 21.35 & 22.07 & 21.88 &   -    &   -    &  507 &  1.37 \\
 82 & 203.36013 & 23.73543 & 0.04721 &  21.22 & 20.19 & 21.36 & 20.98 &   21.83 &   -    & 1701 &  5.45 \\
 88 & 206.98336 &  7.92560 & 0.04369 &  20.18 & 19.12 & 20.22 & 19.72 &   20.15 &   20.29 & 1544 &  3.64 \\
 90 & 210.18102 & 19.85603 & 0.05325 &  21.28 & 20.22 & 21.20 & 20.75 &   21.40 &    -    & 1257 &  1.56 \\
 97 & 223.21934 &  9.47723 & 0.05203 &   -    & 21.23 & 21.95 & 21.91 &   21.94 &   -    &  534 &  3.00 \\
103 & 233.97741 &  7.16452 & 0.04183 &   -    & 21.08 & 22.02 & 21.78 &   22.02 &    -    & 1019 &  0.88 \\
104 & 235.51594 & 22.65047 & 0.04633 &  22.01 & 21.10 & 22.08 & 21.63 &   21.79 &   -    &  629 &  4.34 \\
106 & 240.55233 & 14.76420 & 0.03634 &  20.40 & 19.19 & 20.23 & 20.32 &   20.23 &   20.51 & 1490 &  0.75 \\
108 & 243.45135 & 36.37016 & 0.03892 &  20.98 & 19.73 & 20.94 & 21.01 &   21.57 &   21.02 & 1367 &  4.68 \\
121 & 348.48918 & 29.58686 & 0.04666 &  22.15 & 21.12 & 22.00 & 21.75 &   21.11 &   -    &  422 & 13.37 \\
122 & 350.13252 & 12.42720 & 0.04192 &  21.00 & 20.20 & 21.14 & 21.01 &   21.33 &    -    &  777 &  1.55 \\
\enddata
\tablecomments{(5-8) SDSS $ugri$ magnitudes. (9-10) Pan-STARRS1 $zy$ magnitudes. (12) 3D distances to the nearest galaxy neighbor in SDSS DR12 spectroscopic catalog. The online machine-readable table includes many columns not shown here.}
\end{deluxetable*}


\begin{deluxetable*}{cccccccccccc}
\tablecaption{Properties of blueberry galaxies}
\tabletypesize{\scriptsize}
\centering
\tablehead{\colhead{objID} & \colhead{[OII]3727} & \colhead{[OII]3727$_{err}$} & \colhead{[OIII]4363} &  \colhead{[OIII]4363$_{err}$} & \colhead{[OIII]4959} & \colhead{[OIII]5007}  &\colhead{\hb} & \colhead{\ha} & \colhead{log(M/$M\odot$)} & \colhead{SFR[$M\odot~yr^{-1}$]}  &\colhead{12+log(O/H)} \\
\colhead{(1)} & \colhead{(2)} & \colhead{(3)} & \colhead{(4)} & \colhead{(5)} & \colhead{(6)} & \colhead{(7)} & \colhead{(8)}  & \colhead{(9)} & \colhead{(10)} & \colhead{(11)} & \colhead{(12)} } 
\startdata
  2 &    8.1 &    0.15 &   6.0 &  0.06 &   56 &  172 &  28 &   63 & 6.6 & 0.15 &   7.53 \\
  3 &   51.0 &    0.91 &  13.8 &  0.58 &  199 &  614 &  98 &  310 & 7.2 & 0.50 &   7.79 \\
  4 &    8.5 &    0.10 &   6.6 &  0.13 &   81 &  243 &  35 &  103 & 7.2 & 0.23 &   7.72 \\
  5 &   42.0 &    0.76 &  27.6 &  0.21 &  285 &  867 & 126 &  150 & 8.0 & 0.88 &   7.64 \\
  6 &   -    &    -    &  39.6 &  1.97 &  532 & 1605 & 212 &  666 & 7.1 & 0.37 &   -    \\
  7 &   27.6 &    1.08 &  16.8 &  0.26 &  143 &  422 &  78 &  225 & 6.8 & 0.22 &   7.45 \\
  9 &   26.6 &    0.23 &  10.2 &  0.03 &  147 &  423 &  57 &  174 & 7.3 & 0.31 &   7.83 \\
 10 &  188.9 &    6.65 &  60.9 &  3.84 &  979 & 2961 & 428 & 1341 & 8.1 & 1.61 &   7.87 \\
 12 &   73.8 &    0.66 &  14.6 &  2.60 &  232 &  710 & 117 &  380 & 7.9 & 1.08 &   7.83 \\
 13 &   27.3 &    0.58 &  25.4 &  1.59 &  326 &  978 & 118 &  371 & 7.5 & 0.45 &   7.82 \\
 27 &   65.4 &    0.10 &  21.2 &  0.08 &  258 &  770 & 146 &  211 & 7.7 & 1.16 &   7.62 \\
 28 &   18.9 &    0.29 &   8.7 &  0.15 &   61 &  183 &  38 &   77 & 6.8 & 0.21 &   7.35 \\
 31 &   31.4 &    0.14 &   7.3 &  0.14 &   87 &  261 &  48 &   58 & 7.5 & 0.43 &   7.65 \\
 34 &   37.6 &    1.17 &  23.5 &  0.33 &  279 &  864 & 107 &  216 & 8.3 & 0.54 &   7.78 \\
 38 &    6.1 &    0.05 &   2.6 &  0.01 &   33 &  101 &  16 &   28 & 6.9 & 0.20 &   7.73 \\
 39 &   19.3 &    0.17 &   8.5 &  0.09 &   94 &  289 &  51 &  108 & 7.2 & 0.35 &   7.61 \\
 41 &   17.4 &    0.09 &  16.9 &  0.13 &  233 &  707 &  81 &  174 & 7.3 & 0.49 &   7.88 \\
 43 &   15.5 &    0.09 &   3.6 &  0.04 &   46 &  138 &  31 &   59 & 6.7 & 0.21 &   7.60 \\
 47 &   22.0 &    0.04 &   5.6 &  0.14 &   65 &  209 &  34 &   68 & 6.9 & 0.19 &   7.69 \\
 50 &   26.5 &    0.05 &   6.7 &  0.05 &   92 &  277 &  46 &   88 & 7.0 & 0.45 &   7.74 \\
 54 &   31.1 &    0.05 &   4.0 &  0.05 &   61 &  190 &  34 &   71 & 7.1 & 0.22 &   7.81 \\
 57 &   13.3 &    0.07 &   3.6 &  0.10 &   54 &  166 &  25 &   51 & 7.6 & 0.14 &   7.82 \\
 59 &    3.9 &    1.78 &   7.4 &  0.08 &   54 &  165 &  50 &   95 & 6.6 & 0.44 &   7.13 \\
 66 &   42.0 &    0.06 &  20.8 &  0.09 &  269 &  819 & 113 &  254 & 6.8 & 0.22 &   7.78 \\
 67 &  277.2 &    0.24 &  49.2 &  0.90 &  954 & 2583 & 386 &  916 & 8.6 & 1.48 &   7.94 \\
 68 &   22.4 &    0.07 &  12.5 &  0.11 &  136 &  414 &  65 &  118 & 7.2 & 0.58 &   7.64 \\
 69 &  -     &    -    & 107.2 &  0.95 &   25 &   73 &  11 &   24 & 6.7 & 0.05 &    -   \\
 70 &   30.5 &    0.03 &   9.9 &  0.04 &  113 &  339 &  53 &  114 & 8.0 & 0.34 &   7.67 \\
 75 &   12.1 &    0.04 &   4.9 &  0.04 &   56 &  166 &  27 &   48 & 6.8 & 0.26 &   7.65 \\
 78 &    4.7 &    0.04 &   1.2 &  0.02 &   13 &   42 &   7 &   12 & 6.6 & 0.06 &   7.64 \\
 82 &   12.3 &    0.60 &   5.8 &  0.09 &   58 &  206 &  28 &   50 & 6.7 & 0.23 &   7.71 \\
 88 &   51.2 &    0.07 &   9.9 &  0.20 &  184 &  564 &  78 &  168 & 8.5 & 0.54 &   7.98 \\
 90 &    9.3 &    0.04 &   5.2 &  0.02 &   68 &  200 &  22 &   45 & 7.1 & 0.24 &   7.86 \\
 97 &    7.2 &    0.15 &   1.4 &  0.05 &   17 &   51 &   9 &   17 & 6.7 & 0.09 &   7.70 \\
103 &    9.2 &    0.15 &   1.8 &  0.11 &   26 &   82 &  13 &   28 & 6.6 & 0.09 &   7.81 \\
104 &    8.9 &    0.11 &   1.5 &  0.10 &   18 &   63 &  14 &   25 & 6.7 & 0.11 &   7.60 \\
106 &   37.1 &    9.34 &   7.9 &  1.14 &  205 &  611 &  77 &  324 & 8.0 & 0.36 &   -  \\
108 &    4.6 &    2.21 &  10.2 &  0.05 &   93 &  291 &  41 &   76 & 6.6 & 0.23 &   7.57 \\
121 &    8.8 &    0.21 &   2.4 &  0.50 &   16 &   52 &  19 &   24 & 6.7 & 0.15 &   7.17 \\
122 &   16.8 &    0.41 &   4.5 &  0.08 &   56 &  167 &  32 &   67 & 6.8 & 0.21 &   7.64 \\
\enddata
\tablecomments{(2-9) Line fluxes and errors are in units of $10^{-16}~erg~s^{-1}~cm^{-2}$; (10) Stellar mass from fitting the $ugrizy$ photometric data; (11) Star formation rate from \ha\ emission line; (12) metallicity measured with the [OIII]4363 line. The online machine-readable table includes many columns of line measurements not shown here.}
\end{deluxetable*}

\clearpage

\acknowledgments
We thank Chun Ly, Lin Lin, Chenwei Yang, and Tianxing Jiang for help with the data reduction and observations. Observations reported here were obtained at the MMT Observatory, a joint facility of the University of Arizona and the Smithsonian Institution. H.Y. and J.X.W. thank support from NSFC 11233002, 11421303, and CAS Frontier Science Key Research Program (QYZDJ-SSW-SLH006). This work has also been supported in part by NSF grant AST-1518057; and by support for HST program \#14201, which was provided by NASA through a grant from the Space Telescope Science Institute, which is operated by the Association of Universities for Research in Astronomy, Inc., under NASA contract NAS 5-26555.

\appendix

\section{Selection Criteria in SDSS DR12 Photometric Catalog}
We use the following criteria to select the photometric candidates in the SDSS DR12 CasJobs tool. To make the selection robust, we used both the cModelMag and psfMag. We also require the Galactic latitude $>20$ or $<-20$ degree. The ``calibStatus" flag is the status of photometric calibration. 

\begin{lstlisting}[basicstyle=\ttfamily]
SELECT objID, ra, dec   into blueberry   from PhotoObj
WHERE cModelMag_g>0  and  cModelMag_g<23  and  psfMag_g<23
  and cModelMagErr_g<0.2 and  cModelMagErr_r<0.2
  and cModelMag_u - cModelMag_g >0.3  and  psfMag_u - psfMag_g >0.3
  and cModelMag_r - cModelMag_g >0.5  and  psfMag_r - psfMag_g >0.3
  and cModelMag_r - cModelMag_i <1.0  and  psfMag_r - psfMag_i <1.0
  and (cModelMag_r - cModelMag_g >0.7  or  cModelMag_i - cModelMag_g >1.0)
  and (psfMag_r - psfMag_g >0.7  or  psfMag_i - psfMag_g >1.0)
  and cModelMag_i - cModelMag_g >0.5
  and petroR90_r<5.0  and  petroR90_r>0  and  petrorad_g<5.0  -- radius
  and (b>20 or b<-20)   -- Galactic latitude
  and calibStatus_u <=2  -- Photometric observation status
  and calibStatus_g <=2  and calibStatus_r <=2  and calibStatus_i <=2
  and clean=1 -- "clean" photometry 
  and (flags & 0x0101000980000010)=0  -- not MOVED, BAD_MOVING_FIT, 
     --  DEBLENDED_AS_MOVING, CHILD, MAYBE_CR, TOO_FEW_GOOD_DETECTIONS
\end{lstlisting}

\end{document}